\newcommand{\be}{\begin{equation}}
        \newcommand{\ee}{\end{equation}}
        \newcommand{\ba}{\begin{eqnarray}}
        \newcommand{\ea}{\end{eqnarray}}
        \newcommand{\ban}{\begin{eqnarray*}}
        \newcommand{\ean}{\end{eqnarray*}}
        \newcommand{\nl}{\nonumber \\} 
  
\newcommand{\C}{{\mathbb C}}

\newtheorem{theorem}{Theorem} 
\newtheorem{lemma}{Lemma} 
\newtheorem{corollary}{Corollary} 

\documentclass {article} 
\usepackage {graphicx,amsfonts,latexsym}

\begin{document}

      \begin{center}
      {\Large \bfseries 
      Continuum spin foam model for $3$d gravity
      \\}
      \vspace{0.4cm}
      {\em 
      Jos\'e A. Zapata
      \\}
      \vspace{0.2cm}
      {\small 
      Instituto de Matem\'aticas UNAM \\
      Morelia Mich. 58090 M\'exico \\
      {\small \ttfamily 
      zapata@math.unam.mx
      \\ }}
      \end{center}
\begin{abstract} 
An example illustrating a continuum spin foam framework is presented. 
This covariant framework induces the  kinematics of canonical loop 
quantization, and its dynamics is generated by a {\em renormalized} sum 
over colored polyhedra. 

Physically the example corresponds to $3$d gravity with cosmological
constant.  Starting from a kinematical structure that
accommodates local degrees of freedom and does not involve the choice 
of any background structure (e. g. triangulation), the dynamics
reduces the field theory to have only global degrees of freedom.  
The result is {\em projectively} equivalent to the Turaev-Viro model. 
\end{abstract}





\section*{\large I. INTRODUCTION}

Several TQFT's can be written as a sum over assignments of 
spins to polyhedra: that is, as spin foam models. 
A trend in today's research is to try to find a model of a similar 
type that is related to $4$d gravity \cite{SF}. 
Regarding these models as fundamental
amounts to postulating that gravity is only effectively a field theory, but 
fundamentally it has only finitely many degrees of freedom and
a privileged polyhedron (or triangulation) 
comes along with spacetime. 
It has been proposed to get rid of this extra structure with a sum 
over triangulations \cite{SumTriang}. 
In this article we will explore another route. 
We will regard the continuum as fundamental and 
take the diffeomorphism invariance of general relativity 
as the guiding symmetry. 
We share this principle with canonical loop quantization \cite{lqg}.


In this paper we present an example. It is defined in the 
continuum, but it turns out to be equivalent to the Turaev-Viro model 
\cite{T-V}. 
The interest of our example relies on the fact that it is a continuum 
spin foam model. More precisely, the construction induces the 
kinematics of q-deformed loop quantization 
in the spatial slices, and the projector to the 
space of physical states is constructed as a renormalized 
sum over colored (by spins) polyhedra. 

The construction does not involve the choice of any background 
structure, and the diffeomorphism group acts faithfully at the 
kinematical level. 

The strategy used to generate this family of examples (one for every 
value of a deformation parameter) can be adapted to 
other spin foam models. A ``renormalizability 
condition'' determines whether the continuum theory exists. The proof 
that other topological models satisfy the condition proceeds almost in 
complete parallel to the proof given here. Interestingly, there are 
reasons to believe that there are other examples corresponding to 
genuine field theories. In the case of compact QED, 
the first steps in this direction have already been taken \cite{CQED}. 

In section II we construct the example and prove the 
equivalence with the Turaev-Viro model. Section III gives an 
interpretation of the projector as a renormalized sum over quantum 
geometries.

\section*{\large II. A CONTINUUM SPIN FOAM MODEL}
This is the central section of the paper. In the beginning 
we present the construction of the example and show its main 
properties. With minor modification of the proofs, the construction 
applies to other topological spin foam models and the same strategy 
could apply to other less trivial spin foam models as well.  In the last 
subsection, 
we prove the projective equivalence with the Turaev-Viro model. 


\subsection*{\large A. Embedded graphs and embedded polyhedra}

In this work all the spaces and maps are piecewise linear. 

Consider a compact surface without boundary $\Sigma$. 
By an embedded graph $\Gamma$ we mean a finite one-dimensional 
$CW$-complex all whose vertices have valence two or bigger, 
together with an embedding into $\Sigma$. 
The set of all graphs embedded into a given surface will be denoted 
by $G(\Sigma)$. This set has a natural partial order given by inclusion 
and it is directed. 

Similarly, consider a $3$-manifold with boundary $M$. 
By an embedded polyhedron $X$ we mean a finite two-dimensional 
$CW$-complex all whose internal edges have valence two or bigger, 
together with an embedding into $M$ such that 
$X\cap \partial M$ denoted by $\partial X$ belongs to 
$G(\partial M)$. 
The set of all polyhedra embedded into a given $3$-manifold will be 
denoted by $P(M)$. Again, this set is partially ordered and directed.

\subsection*{\large B. Data from lattice gauge theory}

Our construction starts with data generated by ``lattice gauge theory'' 
on all lattices embedded into given spacetimes. 

For every $\Gamma \in G(\Sigma)$ we are given a complex vector 
space $C(\Gamma)$ and for every $X \in P(M) $ we are given a 
linear functional 
$\Omega^n_{X}е: C(\partial X) \to {\rm \bf C}$. Thus, for each surface 
$\Sigma$ we get a collection of vector spaces labeled by 
embedded graphs 
and for each $3$-manifold $M$ we get a collection of linear 
functionals labeled 
by embedded polyhedra. Our work is based on the compatibility of these 
collections of structures with the partial order of the labeling sets. 

Now we define these objects in the example presented here. 
We follow the notation of Turaev and Viro \cite{T-V} as closely as 
possible as well as their conventions and normalizations. 

Fix an integer $r\geq 3$ and denote by $I$ the set of spins 
$\{ 0, 1/2, \ldots , (r-2)/2 \}$. 
A triple of spins is admissible, $(i, j, k) \in adm$, if $i +j+k$ is an 
integer and 
\[
i \leq j+k , \quad j \leq k+i , \quad k \leq i+j , \quad i+j+k \leq 
r-2. 
\]

Given an embedded graph $\Gamma \in G(\Sigma)$ we assign to it an 
abstract simple graph $\Gamma_{s}е$. The vertices of a simple graph 
are always trivalent; a graph dual to a 
triangulation is an example of a simple graph. 

\begin{figure}[ht]
\centering
\includegraphics[scale=0.75]{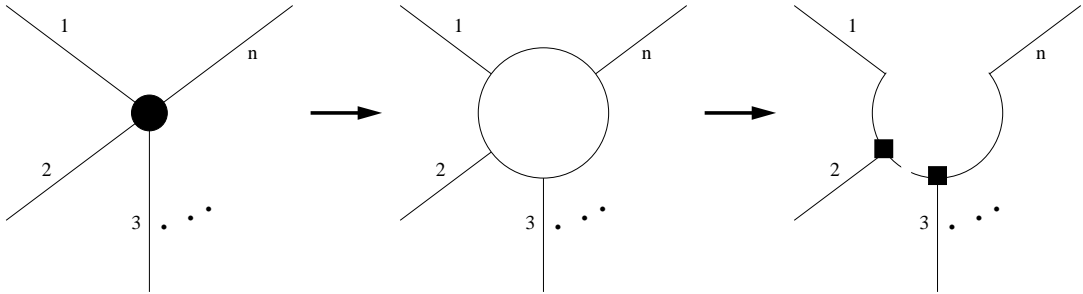}
\vspace*{7pt}
\caption{
The construction uses an arbitrary numbering of the 
edges coming to a vertex. Internal vertices are shown as  
square dots and internal edges are dashed lines. 
}
\label{fig1}
\end{figure}

$\Gamma_{s}е$ is constructed by blowing up the vertices of 
$\Gamma$ as we explain below. To each edge of $\Gamma$ 
corresponds one edge in $\Gamma_{s}е$ and to each vertex of 
$\Gamma$ we assign several ``internal vertices'' and ``internal 
edges'' in $\Gamma_{s}е$ in the manner indicated by Figure 
\ref{fig1}.

\setlength{\unitlength}{1cm}

A coloring $\alpha$ of $\Gamma$ is an assignment of spins 
to its edges and intertwiners to its vertices. An intertwiner is a 
coloring (with spins) of the internal edges assigned to the vertex. 
Clearly, a coloring $\alpha$ of $\Gamma$ induces a coloring of 
$\Gamma_{s}е$, an assignment of spins to its edges. 
The coloring is admissible, $\alpha \in adm(\Gamma)$, if 
at each vertex $v$ of $\Gamma_{s}е$ the triple 
$(\alpha(e_{1}е(v)), \alpha(e_{2}е(v)), \alpha(e_{3}е(v))) \in adm$. 

Then $C(\Gamma)$ is defined by 
\[
C(\Gamma) = \C[adm(\Gamma)] 
\]
where $\C[adm(\Gamma)]$ is the complex vector space freely 
generated by the elements of $adm(\Gamma)$; in addition, 
the inner product makes $adm(\Gamma)$ an orthonormal set. 
A different choice of internal structure in the construction of 
$\Gamma_{s}е$ results in an a priori different vector space; however 
``recoupling moves'' on the internal edges define an isomorphism 
between any two spaces generated by different choices. This is 
reviewed in the Appendix. We identify all these isomorphic vector spaces 
and call them $C(\Gamma)$. 

In the non q-deformed case, when the set of spins is infinite, the 
space just constructed is the space of square integrable functions 
of (generalized) $SU(2)$-connections \cite{A-L}, and it is the heart 
of the kinematics of canonical loop quantum gravity \cite{lqg}.

$X_{s}е$, a simple abstract polyhedron,  
is constructed by blowing up the edges and vertices 
of $X\in P(M)$ as we explain below. 
We proceed in complete parallel to the case of graphs. 
First, we surround every edge by a cylindrical neighborhood and every 
vertex by a spherical neighborhood; 
in this way we created an internal (empty) 
bubble for each edge and vertex of $X$ in analogy with the middle 
picture of Figure \ref{fig1}. Then, we will erase 
one internal face in each internal bubble, as we did in the last 
picture of Figure \ref{fig1}, to create $X_{s}е$. 
There are only two rules to select the face to be erased from each 
internal bubble. In the case of edge-bubbles, we erase one of the 
lateral faces (not a face shared with a vertex-bubble). And in the 
case of the vertex-bubbles, we erase one of the faces shared with an 
edge-bubble. Notice that the valence of any edge in $X_{s}е$ is three 
and that at every vertex of $X_{s}е$ six faces meet. 

A coloring $\varphi$ of $X$ is an assignment of spins 
to its faces and intertwiners to its edges. An intertwiner is a 
coloring (with spins) of the internal faces assigned to the edge. 
The coloring is admissible, $\varphi \in adm(X)$, if 
at each edge $e$ of $X_{s}е$ the triple 
$(\varphi(f_{1}е(e)), \varphi(f_{2}е(e)), \varphi(f_{3}е(e))) \in adm$. 

Each colored polyhedra is assigned a weight. This assignment is a 
simple extension of 
the Turaev and Viro weight to embedded (not necessarily 
simple) polyhedra. The Turaev-Viro weight is constructed as the product 
of weights assigned to its vertices and faces (with a correction due 
to the boundary edges) \cite{T-V}. 
\[
 | X |_{\varphi} =  | X _{s} |^{TV}_{\varphi} = 
 w^{-2\chi(X_{s}е)+\chi(\partial X_{s}е)}
 \prod_{f\in F(X_{s}е)}еw_{\varphi(f)}^{2\chi(f)е}  
 \prod_{e\in E(X_{s}е)}еw_{\partial \varphi(e)}е^{\chi(e)} 
 \prod_{v\in V(X_{s}е)} | \hat{T}_{v} |_{\varphi}ее
\]
Here $\chi$ denotes Euler characteristic. 
$w_{\varphi(f)}е$ and $w_{\partial \varphi(e)}$ denote the 
quantum group analog of the 
dimension of the spin $j$ representation for 
$j= \varphi(f)$ and $j= \partial \varphi(e)$ respectively. 
$| \hat{T}_{v} |_{\varphi}$ is the quantum $6j$ symbol corresponding 
to the $6$-tuple of spins of the faces meeting at vertex $v$. 

Then $\Omega_{X}е: C(\partial(X)) \to \C$ is defined by 
$\Omega_{X}е(\alpha) = \sum_{\varphi}е | X |_{\varphi}е$, 
where the sum runs over the colorings $\varphi \in adm(X)$ that 
induce $\alpha$ in the boundary. We will work with the 
linear functional normalized dividing by the ``vacuum to vacuum 
amplitude'' $\Omega_{X}е(j=0)$, 
\[
\Omega^n_{X}е(\alpha) = 
\frac{\Omega_{X}е(\alpha)}{\Omega_{X}е(j=0)} .
\]
Note that $\Omega^n_{X}е(\alpha) $ can also be defined from 
$\Omega'_{X}е(\alpha) = \sum_{\varphi}е | X |'_{\varphi}е$ which is 
defined from a less refined 
Turaev-Viro weight $| X |'_{\varphi}е$ in which the normalizing factor 
$w^{-2\chi(X_{s}е)+\chi(\partial X)}$ is omitted. 

The construction of $X_{s}е$ involves a choice of ``internal 
structure'' on the edges of $X$. However, the invariance 
results of Turaev and 
Viro imply that $\Omega_{X}$ and $\Omega'_{X}$ are independent of 
this choice. This is reviewed in the Appendix.

The weight assigned to a colored polyhedron is what defines the 
dynamics of this example. We extended the Turaev-Viro weight in a 
simple manner, but it is possible to derive a weight assignment for 
general polyhedra that reduces to the Turaev-Viro weight in the case 
of simple polyhedra \cite{girelli+}.

\subsection*{\large C. From lattices to the continuum}

Now we will leave single ``lattices'' and go to the continuum. 

Consider two embedded graphs $\Gamma_{1}е \leq \Gamma_{2}е$ in 
$G(\Sigma)$. It is easy to see that $\alpha \in adm(\Gamma_{1})$ 
defines an admissible coloring in $\Gamma_{2}е$ by 
simply extending the 
coloring with color $j=0$ in the additional edges and extending the 
intertwiners also coloring with $j=0$ in the additional internal 
edges. Thus, we can take this natural inclusion for granted and write 
$C(\Gamma_{1}е) \subset C(\Gamma_{2}е)$. Due to this property we can 
define $C(\Sigma)$ as the inductive limit or co-limit \cite{footnote} 
of the nested spaces labeled by graphs, 
\[
C(\Sigma) = \hbox{co-}\!\!\lim_{\Gamma \to \Sigma}еC(\Gamma) .
\]

Similarly, for $\alphaе\in C(\partial M)$ induced by 
$\alpha  \in adm(\Gamma)$ we define 
\be
\Omega^n_{M}е(\alphaе) = 
\lim_{X \to M} \Omega^n_{X} (\alpha) , \quad \partial X\geq \Gamma . 
\label{ren}
\ee
When the limit exists it defines 
$\Omega^n_{M}е: C(\partial M) \to C$. 
In our example we prove its existence in the next subsection. 
However, the analogous limit in theories constructed 
from other lattice gauge theories may not exist. The existence of the 
limit should be interpreted as a renormalizability condition. 

Note that the extendibility of colorings from subgraphs 
implies that 
$adm(\Sigma)$ and similarly $adm(M)$ can be defined. A coloring in 
$\alpha \in adm(\Sigma)$ should be thought of as a coloring of all 
graphs, bigger than a certain minimal graph $\Gamma(\alpha)$, 
that is compatible with inclusion of graphs. 
We then have the equivalent definition 
$C(\Sigma)= \C [adm(\Sigma)]$.

\subsection*{\large D. Renormalizability and cellular decompositions}

A polyhedron $X \in M$ 
induces a cellular decomposition for 
$(M, \Gamma)$, 
where $\Gamma \in G(\partial M)$,  if $\partial X = \Gamma$ and 
$M - \partial M -X$ is a disjoint union of open balls. 

\begin{theorem}
Given any polyhedron $X \in P(M)$ there is a finer polyhedron 
$X' \in P(M)$, $X \leq X'$, such that $X'$ 
induces a cellular decomposition for 
$(M, \Gamma)$ for some $\Gamma \geq \partial X$. 
\end{theorem}
{\em Proof.} 
First note that for every manifold with boundary 
one can find embedded polyhedra inducing cellular decompositions. 
Take for example the cellular complex 
dual to any triangulation of the manifold. 
Once we have one 
cellular decomposition, 
we can generate many by subdivision  of 
the induced cells adding faces to the original 
polyhedron. 

Let $X'' \in P(M)$ 
induce a cellular decomposition that is sufficiently fine 
in the sense that for each open ball in 
$\cup_{r=1}е^nB_{r}е = M - \partial M - X''$ 
we have that $B_{r}е - X$ is a finite union of open balls, 
$B_{r}е - X = \cup_{s=1}е^m_{r}еB'_{r,s}е$. 

Since we are working with piecewise-linear spaces, 
such a $X''$ can be constructed from an initial 
cellular decomposition 
after finitely many refinements. 

Let $X' = X'' \cup X$. Then 
$M - \partial M - X' = 
\cup_{r=1}е^n \cup_{s=1}е^m_{r}еB'_{r,s}е$. 
$\Box$

\begin{theorem}\label{T2}
Fix $X \in P(M)$ a 
polyhedron inducing a cellular decomposition 
of $(M, \Gamma)$. 
For any finer polyhedron $X' \in P(M)$,  $X \leq X'$, we have 
\[
\Omega^n_{X} = \Omega^n_{X'}|_{C(\partial X)}е . 
\]
\end{theorem}
The proof of this theorem requires some previous definitions and 
lemmas presented after the following corollary. 

\begin{corollary}[Renormalizability]
In the definition 
\[
\Omega^n_{M}е = \lim_{X \to M} \Omega^n_{X}  
\]
the limit exists. In this sense, the theory defined by the collection 
of linear functionals $\Omega'_{X}$ is renormalizable. 
\end{corollary}
We should mention that for the linear functionals 
$\Omega'_{X}$ and $\Omega_{X}$ the limit does not exist. Only 
the renormalized functional exists in the continuum.

\begin{figure}[ht]
\centering
\includegraphics[scale=0.7]{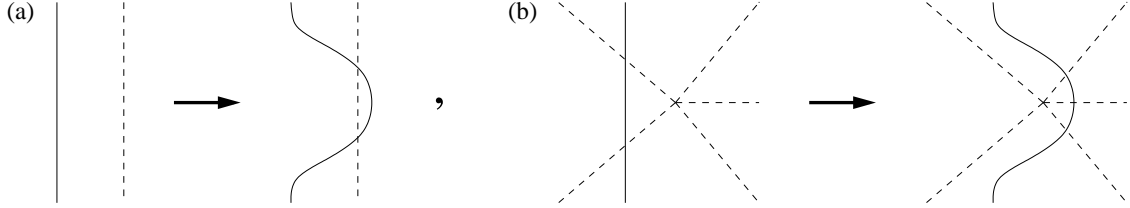}
\vspace*{7pt}
\caption{
In each picture the paper corresponds to a set of faces of a 
portion of the polyhedron. The solid lines indicate edges where one 
(or more) faces meet the paper from above. The dashed lines 
correspond to faces that are below the paper. 
(a) Illustrates a top face(s) sliding through a wedge with bottom 
face(s). 
(b) Illustrates a top face(s) sliding through a corner with bottom faces. 
}
\label{fig2}
\end{figure}

Wedge and corner moves generalize the lune $\cal{L}$ 
and Matveev $\cal{M}$ moves (and their inverses) on simple polyhedra. 
Wedge moves describe a face of an embedded polyhedron $X\in P(M)$ 
sliding through a wedge, while in corner moves a 
face slides through a corner. See Figure \ref{fig2}. 

\setlength{\unitlength}{1cm}

\begin{lemma}[Invariance under wedge and corner moves]\label{moves}
Let $X \in P(M)$ and $\alpha \in adm(\partial X)$. Then 
$\Omega'_{X}е(\alpha)$ is invariant under wedge and corner moves. 
\end{lemma}
{\em Proof.} 
It is clear that a wedge move in $X$ induces a sequence of 
${\cal{L}}$ or ${\cal{L}}^{-1}$ moves in $X_{s}е$. 
Similarly, a corner move 
in $X$ induces a sequence of 
${\cal{M}}$ and/or ${\cal{M}}^{-1}$ and/or 
${\cal{L}}$ and/or ${\cal{L}}^{-1}$ in $X_{s}е$. 
Thus, this lemma is a direct consequence of the similar 
invariance lemmas of Turaev and Viro \cite{T-V}. $\Box$

The following lemma is a well-known consequence of the definition a 
properties of $|X|'_{\varphi}$; see \cite{T-V,K-L}. 
\begin{lemma}[The color $j=0$ is invisible]\label{j=0}
Let $X\in P(M)$ and $\varphi \in adm(X)$. Construct the 
polyhedron $X(\varphi)$ erasing from $X$ the ``invisible'' $2$-strata, 
the ones colored with $j=0$. Then 
\[
|X(\varphi)|'_{\varphi}е= |X|'_{\varphi}е .
\]
Also, for $\alpha \in adm(\partial X)$ construct $X(\alpha)$ erasing 
from $X$ the $2$-strata meeting $\partial X$ in edges colored 
with $j=0$. Then 
\[
\Omega'_{X(\alpha)}е (\alpha)=\Omega'_{X}(\alpha) .
\]
\end{lemma}

{\em Proof of Theorem 2.} 
Consider a collection of nested polyhedra $\{ X_{r}е\}$ such that 
$X=X_{1}\leq X_{2} \leq \ldots \leq X_{n} =X'$ where 
$X_{r+1}е - X_{r}е\subset B_{s(r)}е$ for some cell $B_{s}е$ in 
$\cup_{s=1}^nB_{s}е=M - \partial M - X$ 
with $B_{s(r)}е\not= B_{s(t)}е$ if $r\not= t$. 

We will show that 
$\Omega^n_{X_{r+1}е}|_{C(\partial X_{r}е)}е = \Omega^n_{X_{r}е}$. 

Let $\partial B_{s(r)}-\partial M=\cup_{u=1}^m F_{u}е$ 
with $F_{u}е$ faces of $X$. 

The closures of the 
faces in $X_{r+1}е - X_{r}е$ may intersect several of the faces of 
$\partial B_{s(r)}е$, but 
we can use wedge and corner moves on $X_{r+1}$ (inside $B_{s(r)}е$) 
until only $F_{1}е$ (and $\partial M$ if $B_{s(r)}е$ is a boundary 
cell) is intersected. 
Call the resulting polyhedron $Y'_{r+1}е$.  Remove from $Y'_{r+1}е$ 
the $2$-strata meeting $\partial Y'_{r+1} - \partial X_{r}$; we call 
this polyhedron $Y_{r+1}е$. 

Let $B\subset M$ be an open ball such that 
$Y_{r+1}е - X_{r}е\subset B$, and $B \cap X_{r}е$ lies in the interior 
of $F_{1}е$. 

\begin{figure}[ht]
\centering
\includegraphics[scale=0.85]{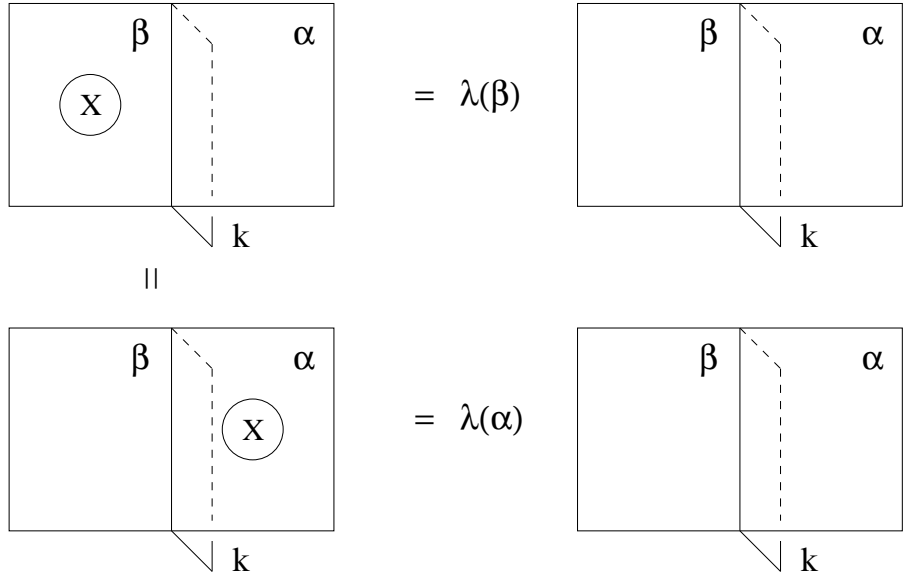}
\vspace*{7pt}
\caption{
The pictures should be thought of as 
$\Omega'$ evaluations with the indicated boundary coloring. 
}
\label{fig3}
\end{figure}

Then 
\ban
\Omega'_{X_{r+1}е}е|_{C(\partial X_{r}е)}е (\alpha) &=& 
\Omega'_{Y_{r+1}е}е(\alpha) = 
\sum_{\beta \in adm(Y_{r+1}е\cap \partial B)}е 
\Omega'_{Y_{r+1}е\cap B^c}е(\alpha \cup \beta) 
\Omega'_{Y_{r+1}е\cap \bar{B}}е(\beta) \\ &=& 
\sum_{\beta \in adm(X_{r}е\cap \partial B)}е 
\Omega'_{X_{r}е\cap B^c}е(\alpha \cup \beta) 
\Omega'_{Y_{r+1}е\cap \bar{B}}е(\beta) \\ &=&
\lambda 
\sum_{\beta \in adm(X_{r}е\cap \partial B)}е 
\Omega'_{X_{r}е\cap B^c}е(\alpha \cup \beta) 
\Omega'_{X_{r}е\cap \bar{B}}е(\beta) = 
\lambda 
\Omega'_{X_{r}}(\alpha)
\ean
where $B^c$ and $\bar{B}$ denote the complement and the closure of 
$B$ respectively. In the first equality 
Lemma \ref{moves} and Lemma \ref{j=0} were used. 
In the second and fifth equalities the associativity of $\Omega'$ was 
used. The fourth equality is due to the following lemma. 
\begin{lemma}
The quotient $\frac{\Omega'_{Y_{r+1}е\cap \bar{B}}е(\beta)}
{\Omega'_{X_{r}е\cap \bar{B}}е(\beta)} = \lambda$ 
is independent of the coloring. 
\end{lemma}
{\em Proof of the Lemma.} 
Let $(\beta, \alpha, k)\in adm$. The proof follows picture in Figure 
\ref{fig3}. 

\setlength{\unitlength}{1cm}

In the upper left picture the face with 
a circle and an $X$ inside represents $Y_{r+1}е\cap \bar{B}$; 
the rest of the faces play only an auxiliary role. Using 
wedge moves the $X$ slides through the edge. The result is the 
picture in the bottom left. There we can again identify the face with 
a circle and an $X$ with $Y_{r+1}е\cap \bar{B}$. From the pictures in 
the left to the ones in the right we have only used 
$\Omega'_{Y_{r+1}е\cap \bar{B}}е(\gamma)
= \lambda(\gamma) \Omega'_{X_{r}е\cap \bar{B}}е(\gamma)$. 

The lemma follows from the fact that the set of spins is 
$adm$-connected; given any two spins there is a chain of admissible 
triples linking them. $\Box$

This completes the proof of Theorem \ref{T2}. $\Box$

\subsection*{\large E. Spaces of physical states and propagators}

We will now define adequate spaces of physical states and 
propagators. To do this cleanly we need 
to use the language of cobordism theory. 
A cobordism $W$ is defined by the triple 
$W=(M; i_{\Sigma_{0}е}е:\Sigma_{0}е \to \partial M, 
i_{\Sigma_{1}е}е:\Sigma_{1}е \to \partial M)$ 
where $M$ is a $3$-manifold with boundary, 
$\Sigma_{0}е$ and $\Sigma_{1}е$ are compact surfaces 
without boundary, 
$\partial M = i_{\Sigma_{0}е}е(\Sigma_{0}) \cup 
i_{\Sigma_{1}е}е(\Sigma_{1})$ and 
$i_{\Sigma_{0}е}е(\Sigma_{0}) \cap 
i_{\Sigma_{1}е}е(\Sigma_{1}) = \emptyset$. 
Cobordisms can be composed in an obvious way. Also, there is a special 
cobordism for each $\Sigma$ compact surface without boundary, 
$id_{\Sigma}е=(\Sigma \times I; i_{0}е(p)= (p, 0), i_{1}е(p)=(p,1))$. 

To every $\Sigma$ we assign a space of physical states 
$H(\Sigma)\subset C(\Sigma)^{*}$ defined by 
$H(\Sigma)= \Omega^n_{\Sigma \times I}(C(\Sigma)) $. 
In the jargon of the field $\Omega^n_{\Sigma \times I}$ is referred to 
as the ``generalized projector,'' even though 
$H(\Sigma) \subset C(\Sigma)^{*}$. 
It is useful to note that 
$H(\Sigma)$ is spanned by vectors $\bar{\alphaе}$ defined by 
\[
\bar{\alphaе} [\delta]
= \Omega^n_{\Sigma \times I}
(i_{0}^{-1*}е\alphaе\cup i_{1}^{-1*}е\delta)
\]
where $\alpha, \delta \in C(\Sigma)$ 
and we have used the pull back maps 
$i_{0}^{-1*}е:adm(\Sigma) \to adm(\partial (\Sigma \times I)_{t=0}е)$, 
$i_{1}^{-1*}е:adm(\Sigma) \to adm(\partial (\Sigma \times I)_{t=1}е)$. 
The inner product in $H(\Sigma)$ is defined by 
\[
(\bar{\alphaе}, \bar{\betaе}) = \bar{\alphaе}(\beta) . 
\]

Now for a general cobordism 
define the map $\Phi^n_{W}: H(\Sigma_{0}) \to C(\Sigma_{1})^{*}$ 
by 
\[
\Phi^n_{W}(\bar{\alpha}) [\delta]= 
\Omega^n_{M}
(i_{0}^{-1*}е\alphaе\cup 
i_{1}^{-1*}е\delta) . 
\]

\begin{theorem}
    $\Phi^n_{W}(H(\Sigma_{0})) \subset H(\Sigma_{1})$
\end{theorem}
{\em Proof.} 
Consider a regular 
neighborhood $M''$ of $i_{\Sigma_{1}е}е\Sigma_{1}е$ in $M$; 
it has the $\Sigma_{1}е \times I$ topology. Denote its boundary by 
$\partial M'' = \Sigma'_{1}е \cup \Sigma_{1}е$. 
Also call $M' = M- M''$; clearly $M' \approx M$ and  
$\partial M'' = \Sigma_{0}е \cup \Sigma'_{1}е$. 

We are gong to compute $\Phi^n_{W}(\bar{\alpha}) [\delta]$ using the 
following auxiliary structures: First, two boundary graphs 
$\Gamma_{0}е, \Gamma_{1}е$ such that 
$\Gamma_{0}е \geq \Gamma(\alpha)$ and 
$\Gamma_{1}е \geq \Gamma(\delta)$. 
Second, $X\in P(M)$ 
inducing a cellular decomposition 
of $(M; i_{\Sigma_{0}}е\Gamma_{0}е \cup i_{\Sigma_{1}}е\Gamma_{1})$ 
such that $X|_{M''} \in P(M'')$ 
induces a cellular decomposition 
of 
$(M''; i_{\Sigma_{0}}е\Gamma_{0}е \cup i_{\Sigma'_{1}}е\Gamma'_{1})$, 
and  $X|_{M'}\in P(M')$ 
induces a cellular decomposition 
of 
$(M'; i_{\Sigma'_{1}}е\Gamma'_{1}е \cup i_{\Sigma_{1}}е\Gamma_{1})$. 

\ba
\Phi^n_{W}(\bar{\alpha}) [\delta]&=&
\frac{\Omega'_{X}(\alpha\cup\delta)}{\Omega'_{X}(j=0)}    = 
\sum_{\beta} \frac{
\Omega'_{X|_{M'}}(\alpha\cup\beta) \Omega'_{X|_{M''}}(\beta\cup\delta)}{
\Omega'_{X}(j=0)} \nl
&=& (\frac{\Omega'_{X|_{M'}}(j=0) \Omega'_{X|_{M''}}(j=0)}{\Omega'_{X}(j=0)})
\sum_{\beta} 
\Omega^n_{M'}(i_{\Sigma_{0}е}^{-1*}е\alpha 
\cup i_{\Sigma'_{1}е}^{-1*}е\beta) 
\Omega^n_{M''}(i_{\Sigma'_{1}е}^{-1*}\beta \cup
i_{\Sigma_{1}е}^{-1*}\delta) \nl
&=&  \lambda(\Gamma'_{1}е)
\sum_{\beta} 
\Omega^n_{M}(i_{\Sigma_{0}е}^{-1*}\alpha 
\cup i_{\Sigma_{1}е}^{-1*}\beta) 
\Omega^n_{\Sigma \times I}(i_{0}^{-1*}\beta \cup
i_{1}^{-1*}\delta)  \label{H1}
\ea
The sums are over $\beta \in adm(\Gamma'_{1}е)$. 
In the last equality we used the invariance under homeomorphisms; 
this property follows from the definitions and 
will be described in the next subsection. We could have used another 
cellular decomposition 
to perform the calculation or another auxiliary surface $\Sigma'_{1}е$. 
The result 
would have been another expression for the same element of $H(\Sigma)$. 

The desired result follows from the above equation.  $\Box$

Note that the maps assigned to identity cobordisms are identity maps, 
$\Phi^n_{id_{\Sigma}е}|_{H(\Sigma)} = id_{H(\Sigma)}$. 
Now we show that $\Phi^n_{W}$ satisfies the projectivized propagator 
condition. 
\begin{theorem}
    For any two composable cobordisms, 
    $W_{1}= (M_{1};  i_{\Sigma_{0}}, i_{\Sigma_{1}})$ and 
    $W_{2}= (M_{2};  i'_{\Sigma_{1}}, i_{\Sigma_{2}})$, we have 
    $\Phi^n_{W_{2}е} \circ \Phi^n_{W_{1}е} = 
    \lambda \Phi^n_{W_{2} \circ W_{1}е}$ . 
    \end{theorem}
{\em Proof.} 
The definition of $\Phi^n_{W}$ requires that its argument is written 
in a canonical form. We use (\ref{H1}). We follow the 
strategy of the previous theorem's proof: with $X$ an auxiliary 
polyhedron giving a cellular decomposition for 
$W_{1}е$ and $Y$ for $W_{2}е$, 
we will write the expression in terms of $\Omega'$ to use 
associativity. Then we will rewrite in terms of the renormalized objects 
of $X \cup Y$, which is  an auxiliary 
polyhedron giving a cellular decomposition for 
$W_{2} \circ W_{1}$. 

\ban
\Phi^n_{W_{2}}\circ\Phi^n_{W_{1}}(\bar{\alpha})
(\delta) &=&  
\lambda(\Gamma'_{1}е)
\sum_{\beta} 
\Omega^n_{M_{1}е}(i_{\Sigma_{0}е}^{-1*}\alpha 
\cup i_{\Sigma_{1}е}^{-1*}\beta) 
\Omega^n_{M_{2}е}({i'}_{\Sigma_{1}е}^{-1*}\beta 
\cup
i_{\Sigma_{2}е}^{-1*}\delta)  \\
 &=&  
\lambda(\Gamma'_{1}е) 
(\frac{\Omega'_{X\cup Y}(j=0)}
{\Omega'_{X}(j=0) \Omega'_{Y}(j=0)})
\Omega^n_{X \cup Y}(\alpha \cup \delta)  \\
 &=&  
\lambda(\Gamma'_{1}е) 
\Lambda(\Gamma_{1}е) 
\Phi^n_{W_{2} \circ W_{1}е}(\bar{\alpha})
(\delta) \qquad \Box 
\ean

Here we proved the last two theorems for the example that we are 
presenting, but the proofs extend with very little change 
to the general case in which the renormalizability condition 
(\ref{ren}) holds.

\subsection*{\large F. Action of the homeomorphism group}

A homeomorphism between two manifolds with boundary 
$f:M \to N$ induces a map among the spaces of embedded 
polyhedra. Colorings are pulled back by this map. Similarly, the 
restriction of the map to the boundaries induces a 
pull back map 
among the spaces of colorings which leads to 
$U_{f}е: C(\partial M) \to C(\partial N)$ defined by the action 
$U_{f}(\alpha)=f^{-1*}(\alpha)е$. 
The following 
identities describe the action of homeomorphisms 
\[
|X|'_{\varphi}е = |f(X)|'_{f^{-1*}е\varphi}е , \quad 
\Omega'_{X}е(\alpha)  = \Omega'_{f(X)}е(f^{-1*}е\alpha) , \quad 
\Omega^n_{M} = \Omega^n_{f(M)} \circ U_{f}  . 
\]
Note that the action of $f$ is not trivial due to its action in the 
boundaries. 
The language of the last subsection already assumes this level of 
invariance; $W$ is 
only sensible to $M$ up to a homeomorphism preserving the boundary. 
By dual action of $U_{f}$ we have a representation of the 
modular group (mapping class group) of $\Sigma$ on $H(\Sigma)$. 
Indeed, if $f$ and $g$ are two isotopic homeomorphisms, 
$U_{f}$ and $U_{g}$ induce the same map in $H(\Sigma)$. In 
particular, if $f$ is connected to the identity 
\[
U_{f}|_{H(\Sigma)} = id_{H(\Sigma)}е . 
\]
Thus, for homeomorphisms connected with the identity we have 
$\Omega^n_{f(M)} = \Omega^n_{M} $. 

In relation with canonical loop quantum gravity we have the following 
question: {\em Is it possible to find first the space of homeomorphism 
invariant states and construct $H(\Sigma)$ from it?} 
The next section gives a satisfactory affirmative answer to this 
question. Here we present a preliminary answer. 

Define the space of states invariant under homeomorphisms connected 
with the identity 
${\cal H}_{hom}е(\Sigma)\subset C(\Sigma)^{*}е$ 
as in \cite{lqg}. This space is spanned by elements of the type
\[
{\tilde \alpha}[\beta]е= 
s(\Gamma(\alpha)) \delta_{[\Gamma(\alpha)][\Gamma(\beta)]}е 
(\alphaе, f_{0}^{-1*}\beta)
\]
where the factor $s(\Gamma(\alpha))$ depends on the symmetry of 
the graph, and 
$f_{0}е$ is a homeomorphism (connected with the identity) on
$\Sigma$ taking $\Gamma(\beta)$ to $\Gamma(\alpha)$. 

Clearly, ${\tilde{\alpha}}= {\tilde{\beta}}$ 
only if there is homeomorphism (connected with the identity) $f_{0}е$ 
such that $U_{f_{0}е}е({\tilde{\beta}})= {\tilde{\alpha}}$. 
Thanks to the invariance properties stated above, we can define 
\be
P({\tilde \alpha})е= {\bar \alpha} .  \label{P}
\ee
In this way the Reisenberger-Rovelli projector, 
$P({\cal H}_{hom}е(\Sigma)) = H(\Sigma)$ \cite{R-R}, reconstructs 
the space of physical states from the space of homeomorphism 
invariant states.

\subsection*{\large G. Projective equivalence with the Turaev-Viro model}

A particularization of the construction of Turaev and Viro \cite{T-V} 
can be described as follows. 
{\em 
To each triangulated surface $T$ assign a the vector space 
$C(T^*)$, where $T^{*}е$ is the graph dual to the triangulation. 
Similarly, to each cobordism between triangulated surfaces $W$ assign 
the map $\Phi_{W}е:C(T^{*}е_{0}е) \to C(T^{*}е_{1}е)$ by the formula 
\[
\Phi_{W}е(\alpha)= \sum_{\beta}е 
\Omega_{\Delta^{*}}е(i_{T_{0}е}^{-1*}е\alpha \cup 
i_{T_{1}е}^{-1*}е\beta) \beta
\]
where $\alpha$ and $\beta$ are admissible colorings of the graphs dual 
to the triangulations $T_{0}е$ and $T_{1}е$ respectively, and 
$\Delta$ is a triangulation of the interpolating manifold of $W$. 

The structures are then refined to construct a TQFT. 
The spaces of physical states are 
$Q(T^{*}е)= C(T^{*}е)/ \ker{\Phi_{id_{T}е}е}$ 
with inner product 
$([\alpha], [\beta])= \Omega_{T \times I}е(\alpha, \beta)$. 
The propagators are induced by $\Phi_{W}$, 
$\Psi_{W}е:Q(T^{*}е_{0}е) \to Q(T^{*}е_{1}е)$, 
$\Psi_{W}е[\alpha] = 
\sum_{\beta}е\Omega_{\Delta^{*}е}е(\alpha, \beta) [\beta]$. 

The invariance result says that two spaces $Q(T)$ constructed from 
different triangulations of the same surface are isomorphic, and that 
the map $\Psi_{W}е:Q(T^{*}е_{0}е) \to Q(T^{*}е_{1}е)$ 
is independent of the 
triangulation up to conjugacy by the mentioned isomorphisms. 
}

Now we compare the Turaev-Viro model with our construction. 

\begin{theorem}
    For any triangulation $T$ of a surface $\Sigma$, 
    the map $\overline{i_{T^{*}}е}: Q(T^{*}е) \to H(\Sigma)$ defined 
    by 
    $\overline{i_{T^{*}е}е}([\alpha])= {\bar{\alpha}}$ 
    is an isomorphism, and for any cobordism $W$ 
    \[
    \overline{i_{T_{1}^{*}е}е}^{-1}\circ\Phi^n_{W}
    \circ\overline{i_{T_{0}^{*}е}е} =
    \Lambda \Psi_{W}е .  
    \]
    \end{theorem}
{\em Proof.} 
Since the dual of any triangulation of a $3$-manifold 
gives a cellular decomposition, 
it is easy to verify that $\overline{i_{T^{*}е}е}$ is a 
dilatation (a multiple of an isometry) by 
$\lambda=
(\Omega_{\Delta_{id_{\Sigma}}^{*}е}(j=0)е)^{-\frac{1}{2}}$ 
(where $\Delta_{id_{\Sigma}}$ 
is a triangulation of $\Sigma \times I$ compatible 
with $T$ in both boundaries), 
and that the map taking $\bar{\alpha}$ to 
$\lambda^2\sum_{\beta}е \bar{\alpha}(\beta) [\beta]$ 
is its inverse. 

$\overline{i_{T_{1}^{*}е}е}^{-1}\circ\Phi^n_{W}
\circ\overline{i_{T_{0}^{*}е}е} =\Lambda \Psi_{W}е$ 
follows from a simple application of definitions. We obtain 
$\Lambda = (\Omega_{\Delta_{id_{\Sigma_{1}}}^{*}е}(j=0)е 
\Omega_{\Delta_{W}^{*}е}(j=0)е)^{-1}$, where 
$\Delta_{W}$ is a triangulation of the interpolating manifold of $W$ 
compatible with $T_{0}$ and $T_{1}$. $\Box$

\section*{\large III. INTERPRETATION AS A RENORMALIZED SUM \\
OVER QUANTUM GEOMETRIES}
\subsection*{\large A. Preliminaries}

Recall that we gave an alternative definition of $C(\Sigma)$ as 
$\C [adm(\Sigma)]$. 
Similarly, after having enunciated the ``color $j=0$ is invisible'' 
lemma, 
we can give an equivalent definition of the linear functional $\Omega'_{X}$. 
\[
\Omega'_{X}е(\alpha) = \sum_{\varphi}е | X(\varphi)|'_{\varphi}е
\]
where the sum runs over colorings $\varphi \in adm(M)$ such that 
$\partial \varphi = \alpha$ and 
$X(\varphi)\leq X$. That is, $X$ serves only to restrict the set of 
colorings allowed in $\Omega'_{X}$. 

In this way, we can see $\Omega^n_{M}е$ of 
formula (\ref{ren}) as a renormalized sum over colorings in which 
the ``restricting box'' $X$ grows infinitely large. 

In particular, {\em the 
``generalized projector'' $\Omega^n_{\Sigma \times I}е$ 
is a renormalized sum over quantum geometries.} 

The term {\em quantum geometry} means a history in 
the form of a colored polyhedron because any slice of it 
defines a quantum geometry in canonical loop quantum gravity. 
In formal path integral quantization of general relativity 
the central object is an 
integral over diffeomorphism classes of metrics. 
Our quantum geometries are not analogous to diffeomorphism 
classes of metrics because they are defined for a 
fixed embedding. {\em Is there an 
interpretation of the ``generalized projector'' as a sum over 
knot classes of colored polyhedra?} Below we give an affirmative answer. 

\subsection*{\large B. Summing over knot-classes of colorings}

Consider $X\in P(M)$, $Y\in [X]$ if and only if there is a 
homeomorphism $f:M \to M$ such that $f|_{\partial M}е= id$ and $f(Y)=X$. 
The set of such classes will be denoted by $KP(M)$. 

There is a natural partial ordering in $KP(M)$ defined by 
$[X]\leq [Y]$ if and only if there are representatives $X\in [X]$, 
$Y\in [Y]$ satisfying $X\leq Y$. It is easy to verify that this 
defines a partial ordering; in addition, $KP(M)$ is directed 
with respect to this partial ordering.

Our goal now will be to use $[X]$ as a restricting 
box for a sum over classes of colorings. Then we will remove the 
restriction for the renormalized sum.

Recall that a homeomorphism $f:M \to M$ acts on admissible colorings 
$\varphi \in adm(M)$ by $f^{-1*}\varphi \in adm(M)$. 
We will denote by $[\varphi]$  the class 
of colorings with respect to the group of homeomorphisms that restrict 
to the identity on $\partial M$. 
Note that $[X(\varphi)]$ is independent of the representative 
$\varphi \in [\varphi]$. 

Since 
$|X(f^{-1*}\varphi)|'_{f^{-1*}\varphi}е =|X(\varphi)|'_{\varphi}е$ 
we can define 
$|[\varphi]|= |X(\varphi)|'_{\varphi}е$ and for 
$\alpha \in adm(\partial M)$ 
\[
\Omega'_{[X]}е(\alpha)= \sum_{[\varphi]}е | [\varphi]|' 
s([X(\varphi)], [X])
\]
where the sum runs over all classes of colorings such that 
$\partial [\varphi]=\alpha$ and $[X(\varphi)]\leq [X]$. The symmetry 
factor $s([X(\varphi)], [X])$ counts in how many distinct ways can 
$[X(\varphi)]$ fit in $[X]$. Then, we have the following 
interpretative result concerning the central object of this work. 
\begin{theorem}
$\Omega^n_{M}$, as defined in equation (\ref{ren}), 
is a renormalized sum over knot classes of colorings. That is, 
\[
\Omega^n_{M}е(\alpha)=lim_{[X] \to M}е\Omega^n_{[X]}е(\alpha) . 
\]
\end{theorem}

The same techniques can be applied using the 
notion of class resulting from 
the group of homeomorphisms that preserves each connected 
component of $\partial M$. The analysis proceeds in complete parallel 
to the one described above, but the resulting linear functional acts 
naturally on ${\cal H}_{hom}$. In addition, 
the space of physical states and the projective propagator are 
naturally isomorphic to the ones constructed here. 
This construction is of interest to loop quantization because 
it gives a construction of the Reisenberger-Rovelli projector 
(\ref{P}) as a renormalized sum over quantum geometries.

\section*{\large Acknowledgments}
The author acknowledges support from CONACyT 34228-E and SNI 21286. 

\section*{\large Appendix: Independence of internal structure}

A choice of internal structure is needed to construct the 
simple graph $\Gamma_{s}е$ from $\Gamma \in G(\Sigma)$. In the next 
paragraphs we will see that different choices lead to naturally 
isomorphic vector spaces $C(\Gamma_{s}е)$ which in 
the main the main body of the paper are denoted collectively by 
$C(\Gamma)$. 

Consider a portion of a graph adjacent to a vertex with valence higher 
than three; call it $\Gamma(v)$.  Use the prescription  of section 
II.B Figure \ref{fig1} 
to generate a simple graph $\Gamma(v)_{s}е$. Fix a coloring $\alpha$ 
of $\Gamma(v)$'s edges, and define 
$C(\Gamma(v)_{s}е, \alpha)= 
\C[adm(\Gamma(v)_{s}е, \alpha)]$ 
as the complex vector space generated by the admissible colorings of 
$\Gamma(v)_{s}е$ that are compatible with $\alpha$. 

Construct $\Gamma(v)^3_{s}е$ from $\Gamma(v)_{s}е$ simply 
by sliding 
``edge 1'' counter clockwise past ``edge 3'' (see Figure \ref{fig1}). 
Continue sliding ``edge 1'' in the same direction to 
generate a sequence of graphs 
$(\Gamma(v)_{s}е, \Gamma(v)^3_{s}е, \Gamma(v)^4_{s}е, \ldots , 
\Gamma(v)^{n-1}_{s}е)$. 
Note that $\Gamma(v)^{n-1}_{s}е$ corresponds to the graph 
$\Gamma(v)_{s}е$ obtained using a different numbering of the edges. 
With this set of moves we can generate all 
the graphs $\Gamma(v)_{s}е$ 
obtained with any counter clockwise (or clockwise) 
oriented numbering of edges. 

Now we are going to show that the spaces 
$C(\Gamma(v)^i_{s}е, \alpha)$, $C(\Gamma(v)^{i+1}_{s}е, \alpha)$ 
are naturally isomorphic. 
After we do so, we will know that 
any two counter clockwise 
(or clockwise) oriented numbering of edges produce different simple 
graphs $\Gamma(v)^a_{s} \not= \Gamma(v)^b_{s}$, but naturally 
isomorphic vector spaces 
$C(\Gamma(v)^a_{s}е, \alpha)\approx C(\Gamma(v)^b_{s}е, \alpha)$. 
Clearly, this implies that different choices of $\Gamma_{s}е$ for 
a given $\Gamma$ lead to naturally isomorphic vector spaces. 

Consider the graph $\Gamma(v)^i_{s}е$ described above.  
Its internal edges can be labeled as 
$(e_{3,4}, e_{4,5}, \\
\ldots , e_{i,1}, e_{1,i+1}, \ldots , e_{n-2, n-1}е)$. Similarly, the 
internal edges of the graph $\Gamma(v)^{i+1}_{s}е$ can be labeled as 
$(e_{3,4}, e_{4,5}, 
\ldots , e_{i,i+1}, e_{i+1,1}, \ldots , e_{n-2, n-1}е)$. 
Using an abbreviated notation for the internal edges, we can write the 
generators of $C(\Gamma(v)i_{s}е, \alpha)$ as 
$(\ldots, j_{i,1}е, j_{1,i+1}е, j_{i+1,i+2}е \ldots)_{i}е$. There is one 
generator for each choice of ``internal spins'' that is compatible 
with the coloring of the external edges $\alpha=(j_{1}е, j_{2}е, 
\ldots, j_{n}е)$. 
In a similar fashion, the 
generators of $C(\Gamma(v)^{i+1}_{s}е, \alpha)$ are denoted by 
$(\ldots, j_{i,i+1}е, j_{i+1,1}е, j_{1,i+2}е \ldots)_{i+1}е$. 
The isomorphism is given by the recoupling move \cite{K-L}, which sends 
the generator 
$(\ldots, j_{i,1}е, j_{1,i+1}е, j_{i+1,i+2}е \ldots)_{i}е$ to 
\[
\sum_{j_{3,1}}е w_{j_{1,i+1}}е w_{j_{i+1,1}}е 
\left| \begin{array}{ccc}
j_{i,1}е& j_{1}е & j_{1,i+1}е \\
j_{1, i+2}е& j_{i+1}е& j_{i+1,1}е
\end{array} \right| 
(\ldots, j_{i,i+1}е, j_{i+1,1}е, j_{1,i+2}е \ldots)_{i+1}е\quad . 
\] 
If we slide back ``edge 1'' to its position in $\Gamma(v)^i_{s}е$ 
the recoupling move gives us another $6j$ symbol. The resulting sum of 
products of two $6j$ symbols is just the orthogonality relation, 
meaning that sliding back ``edge 1'' induces just the inverse 
transformation. This completes the proof; any two choices of internal 
structure for $\Gamma \in G(\Sigma)$ lead to naturally isomorphic 
vector spaces.

Let us now show that $\Omega'_{X}е$ is independent of the 
choice of $X_{s}е$ used to define it. To do it, 
we will construct a sequence of 
lune and Matveev moves that interpolates between any two choices of 
$X_{s}е$ leaving $\Omega'_{X}е$ unchanged. 
We will describe a sequence of moves that changes the 
internal structure of an edge and a sequence that changes the internal 
structure of a vertex. Composing these sequences of moves we can 
generate any of the possible choices of internal structure starting 
from a particular choice. 

We will start describing how to change the internal structure of 
edges. 
First take the case of an edge with at least 
one end in 
the boundary. In this case the internal structure of the edge is fixed 
by the structure of the vertex in the boundary; there is no choice of 
internal structure. (If the edge has two vertices 
in the boundary, they are assumed to have compatible structures). 

Take now the case of an edge that starts and ends at two distinct 
vertices that lie in the interior of $M$. 
We will change the internal structure of the edge sliding ``face 1'' 
in the same spirit used to change the internal structure of a vertex 
in the case of boundary graphs. That is, we will slide ``face 1'' 
around the lateral faces of the 
edge-bubble until it is connected to ``face n''. Iterating this process, 
we can change the location of the hole in the edge-bubble to be any of 
the lateral faces. In the next paragraph we describe a sequence of 
Matveev and lune moves that slides ``face 1'' as we need. 

``Face 1'' meets two vertex bubbles, the bubbles 
corresponding to the vertices at the extremes of the edge. Then we can 
use Matveev moves to slide ``face 1'' past ``face 3'' on the surface 
of both vertex-bubbles. Then we simply pull ``face 1'' through the surface 
of the edge-bubble past ``face 3'' using an inverse lune move. 
We can iterate this process to slide ``face 1'' until it is connected 
to ``face n.'' At this point we have moved the position of 
the hole in the edge-bubble, but we may have not finished our job yet. 
One of the vertex-bubbles may have had its hole 
connecting it to the edge-bubble. 
In this case, the sequence of Matveev moves has made ``face 1'' 
surround all the internal vertices of the vertex bubble. To finish we 
have to slide ``face 1'' on the surface of the vertex-bubble 
past all these vertices using Matveev and inverse Matveev moves. 

Finally, consider the case of an edge that 
closes in itself.  ``Face 1'' can be 
slid using the technique explained above 
for the case in which the edge ended in two internal vertices. 
We have described how to change the internal structure of an edge 
with any location in $M$. In the next paragraph we 
explain how to change the internal structure at vertices. 

Each vertex-bubble has a hole connecting it to an edge-bubble. We may  
want to change the location of the hole in a way that connects the 
vertex-bubble to another edge-bubble. To do it we can simply move the 
face that occupies the place where we want the hole to be to the old 
location of the hole. This moving of the face can be achieved by a 
sequence of Matveev and inverse Matveev moves done inside the 
vertex-bubble. This finishes the construction; any two choices of 
internal structure for $X\in P(M)$ are connected by a sequence of lune 
and Matveev moves (and their inverses). Thus, $\Omega'_{X}е$ is 
independent of any choice.


\begin{thebibliography}{10} 

   \bibitem{SF} J. W. Barrett and L.  Crane, 
``Relativistic spin networks and quantum gravity"  
J. Math. Phys. {\bf 39}, 3296-3302 (1998).\\
J. C. Baez, 
``Spin Foam Models'' 
Class. Quant. Grav. {\bf 15}, 1827-1858 (1998). \\
M. Reisenberger, 
``A lattice worldsheet sum for 4-d Euclidean general relativity'' 
Archive: gr-qc/9711052
      
\bibitem{SumTriang} 
R. De Pietri, L. Freidel, K. Krasnov and C. Rovelli, 
``Barrett-Crane model from a Boulatov-Ooguri field theory over a 
homogeneous space''
Nucl. Phys. {\bf B574}, 785-806 (2000). 

\bibitem{lqg} 
A. Ashtekar, J. Lewandowski, D. Marolf, J. Mourao and  T. Thiemann, 
`` Quantization of diffeomorphism invariant theories 
of connections with local degrees of freedom'' 
J. Math. Phys. {\bf 36},  6456-6493 (1995). \\
J. Baez, 
``Spin networks in gauge theory''
Adv. Math. {\bf 117}, 253-272 (1996). \\
R. De Pietri and C. Rovelli, 
``Geometry eigenvalues and the scalar product from recoupling theory in loop
quantum gravity'' 
Phys. Rev. {\bf D54}, 2664-2690 (1996). \\
J. A. Zapata, 
``A combinatorial approach to diffeomorphism 
invariant quantum gauge theories''
J. Math. Phys. {\bf 38}, 5663-5681 (1997). 

\bibitem{T-V} V. G. Turaev and O. Y. Viro, 
``State sum invariants of 3-manifolds and quantum 6-j symbols" 
Topology {\bf 31}, 865-902 (1992). 

\bibitem{CQED}
M. Reisenberger, 
``Worldsheet formulations of gauge theories and gravity'' 
Archive: gr-qc/9412035. \\
J. M. Aroca, H. Fort and R. Gambini,
``Path Integral for the Loop Representation of Lattice Gauge Theories''
Phys. Rev. {\bf D 54}, 7751-7756 (1996). 

\bibitem{A-L} A. Ashtekar and J. Lewandowski, 
``Projective techniques and functional integration for gauge theories'' 
J. Math. Phys. {\bf 36}, 2170-2191 (1995). 

\bibitem{girelli+} F. Girelli, R. Oeckl and A. Perez, 
``Spin foam diagrammatics and topological invariance'' 
Class. Quant. Grav. {\bf 19}, 1093-1108 (2002). 

\bibitem{K-L} L. H. Kauffman and S. L. Lins, 
{\it Temperley-Lieb recoupling theory and invariants of $3$-manifolds}, 
Ann. of Math. Stud.,134, (Princeton Univ. Press, Princeton, NJ, 1994). 

\bibitem{footnote}
For a  definition of the inductive limit see for example \\
G. K\"othe, 
{\it Topological vector spaces I}, 
Grundlehren der mathematischen Wissenschaften 159, 
(Spinger-Verlag Berlin, Heidelberg 1983). \\
The same limit is used to define the kinematical Hilbert space in 
canonical loop quantization \cite{lqg}. 

\bibitem{R-R} M. P. Reisenberger and C. Rovelli, 
``Sum over surfaces form of loop quantum gravity'' 
Phys. Rev. {\bf D 56}, 3490-3508 (1997). 
    
\end{thebibliography}
\end{document}